# Self-organizing the Abstract: Canvas as a Swarm Habitat for Collective Memory, Perception and Cooperative Creativity·

**Vitorino Ramos**

CVRM-IST, Technical Univ. of Lisbon, Av. Rovisco Pais, 1049-001, Lisbon, PORTUGAL,
http://alfa.ist.utl.pt/~cvrm/staff/vramos, vitorino.ramos@alfa.ist.utl.pt

**Abstract**: Many animals can produce very complex intricate architectures that fulfil numerous functional and adaptive requirements (protection from predators, thermal regulation, substrate of social life and reproductive activities, etc). Among them, social insects are capable of generating amazingly complex functional patterns in space and time, although they have limited individual abilities and their behaviour exhibits some degree of randomness. Among all activities by social insects, nest building, cemetery organization and collective sorting, is undoubtedly the most spectacular, as it demonstrates the greatest difference between individual and collective levels. Trying to answer how insects in a colony coordinate their behaviour in order to build these highly complex architectures, scientists assumed a first hypothesis, anthropomorphism, i.e., individual insects were assumed to possess a representation of the global structure to be produced and to make decisions on the basis of that representation. Nest complexity would then result from the complexity of the insect's behaviour. Insect societies, however, are organized in a way that departs radically from the anthropomorphic model in which there is a direct causal relationship between nest complexity and behavioural complexity. Recent works suggests that a social insect colony is a decentralized system composed of cooperative, autonomous units that are distributed in the environment, exhibit simple probabilistic stimulus-response behaviour, and have only access to local information. According to these studies at least two low-level mechanisms play a role in the building activities of social insects: *Self-organization* and discrete *Stigmergy*, being the latter a kind of indirect and environmental synergy. Based on past and 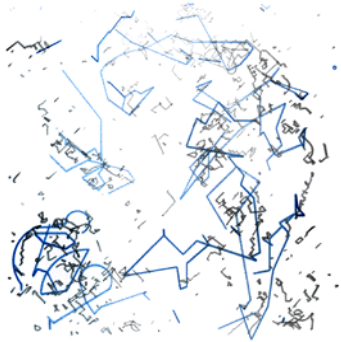 present stigmergic models, and on the underlying scientific research on *Artificial Ant Systems* and *Swarm Intelligence*, while being systems capable of emerging a form of collective intelligence, perception and *Artificial Life*, done by Vitorino Ramos, and on further experiences in collaboration with the plastic artist Leonel Moura, we will show results facing the possibility of considering as "art", as well, the resulting visual expression of these systems. Past experiences under the designation of "Swarm Paintings" conducted in 2001, not only confirmed the possibility of realizing an artificial art (thus non-human), as introduced into the process the questioning of creative migration, specifically from the computer monitors to the canvas via a robotic harm. In more recent self-organized based research we seek to develop and profound the initial ideas by using a swarm of autonomous robots (ARTsBOT project 2002-03), that "live" avoiding the purpose of being merely a simple perpetrator of order streams coming from an external computer, but instead, that actually co-evolve within the canvas space, acting (that is, laying ink) according to simple inner threshold stimulus response functions, reacting simultaneously to the chromatic stimulus present in the canvas environment done by the passage of their team-mates, as well as by the distributed feedback, affecting their future collective behaviour. In parallel, and in what respects to certain types of collective systems, we seek to confirm, in a physically embedded way, that the emergence of order (even as a concept) seems to be found at a lower level of complexity, based on simple and basic interchange of information, and on the local dynamic of parts, who, by self-organizing mechanisms tend to form an lived whole, innovative and adapting, allowing for emergent open-ended creative and distributed production.